\documentclass[
amsmath, amssymb, 
aps, pra,
reprint, twocolumn,
superscriptaddress,
longbibliography,
floatfix,
showkeys
]{revtex4-1}

\usepackage{graphicx, caption}
\usepackage{csquotes}
\usepackage[urlcolor=blue, colorlinks=true,citecolor=blue]{hyperref}

\usepackage{color}
\usepackage{bbm}
\usepackage{nicefrac}
\usepackage{subfig}
\usepackage{cleveref}
\usepackage{float}

\def\bra#1{\ensuremath{\mathinner{\langle{#1}|}}}
\def\ket#1{\ensuremath{\mathinner{|{#1}\rangle}}}

\providecommand{\keywords}[1]
{
  \small	
  \textbf{\textit{Keywords---}} 
}

\begin{document}

\title{Layerwise learning for quantum neural networks}

\author{Andrea Skolik}
    \affiliation{Volkswagen Data:Lab, Ungererstra\ss e 69, 80805 Munich, Germany}
    \affiliation{Ludwig Maximilian University, Theresienstra\ss e 39, 80333 Munich, Germany}
    \affiliation{Leiden University, Niels Bohrweg 1, 2333 CA Leiden, The Netherlands}
\author{Jarrod R. McClean}
    \affiliation{Google Research, 340 Main Street, Venice, CA 90291, USA}
\author{Masoud Mohseni}
    \affiliation{Google Research, 340 Main Street, Venice, CA 90291, USA}
\author{Patrick van der Smagt}
    \affiliation{Volkswagen Group Machine Learning Research Lab, Munich, Germany}
    \affiliation{E\"otv\"os Lor\'and University, Budapest, Hungary}
\author{Martin Leib}
    \affiliation{Volkswagen Data:Lab, Ungererstra\ss e 69, 80805 Munich, Germany}

\begin{abstract}
With the increased focus on quantum circuit learning for near-term applications on quantum devices, in conjunction with unique challenges presented by cost function landscapes of parametrized quantum circuits, strategies for effective training are becoming increasingly important.  In order to ameliorate some of these challenges, we investigate a layerwise learning strategy for parametrized quantum circuits. The circuit depth is incrementally grown during optimization, and only subsets of parameters are updated in each training step. We show that when considering sampling noise, this strategy can help avoid the problem of barren plateaus of the error surface due to the low depth of circuits, low number of parameters trained in one step, and larger magnitude of gradients compared to training the full circuit. These properties make our algorithm preferable for execution on noisy intermediate-scale quantum devices. We demonstrate our approach on an image-classification task on handwritten digits, and show that layerwise learning attains an 8\% lower generalization error on average in comparison to standard learning schemes for training quantum circuits of the same size. Additionally, the percentage of runs that reach lower test errors is up to 40\% larger compared to training the full circuit, which is susceptible to creeping onto a plateau during training.
\end{abstract}

\keywords{Quantum neural network, parametrized quantum circuit, quantum machine learning, gate model quantum computing}

\maketitle

\section{Introduction}
\label{sec:introduction}
Parametrized quantum circuits (PQC) with variational objectives have been proposed as one of the candidate methods to achieve quantum advantage in the era of noisy intermediate-scale quantum (NISQ) devices. Unlike fully error corrected quantum computers, NISQ devices are already available and exceed the performance of classical computers on specific computational tasks~\cite{arute2019quantum}.  This makes explorations of this new computing capability for real-world problems especially interesting within the next decade. Due to their versatility, PQCs are applicable in a multitude of areas like machine learning \cite{broughton2020tensorflow,Grant2018,havlivcek2019supervised,Benedetti2019,Liu2018,Romero2017quantum,Carolan2019,BenedettiPerdomo2019}, optimization \cite{Farhi:2014,Verdon2019metalearning,Hadfield2019,wang2018quantum,streif2020training}, and chemistry \cite{Peruzzo:2014,McClean:2016Theory,OMalley:2016,Yung:2014,Wecker2015,Hempel:2018,kandala2017hardware,McClean:2017hybrid,Rubin2018,Shen2017Quantum,Colless:2018,Santagati:2018,Verdon2019metalearning,mcclean2020decoding,takeshita2020increasing}. 

A PQC consists of an arbitrary set of quantum gates, which are parametrized by a set of controls that determine the outcome of the circuit. A specific type of PQC are so-called ``quantum neural networks'' (QNNs), where the suitable set of parameters for a task is learned based on a given set of data, resembling the training of a classical neural network (NN) \cite{farhi18_QNN,McClean2018,chen2018universal,broughton2020tensorflow}. While the gradient-based backpropagation algorithm \cite{Rumelhart1986} is one of the most successful methods used to train NNs today, its direct translation to QNNs has been challenging. For QNNs, parameter updates for minimizing an objective function are calculated by stochastic gradient descent, based on direct evaluation of derivatives of the objective with respect to parameters in a PQC. The PQC is executed on a quantum device, while the parameter optimization routine is handled by a classical outer loop. The outer loop's success depends on the quality of the quantum device's output, which in the case of gradient-based optimizers are the partial derivatives of the loss function with respect to the PQC. As \cite{McClean2018} has shown, gradients of random PQCs vanish exponentially in the number of qubits, as a function of the number of layers. Furthermore,  \cite{cerezo2020cost} shows that this effect also depends heavily on the choice of cost function, where the barren plateau effect is worst for global cost functions like the fidelity between two quantum states. When executed on a NISQ device, the issue is further amplified because small gradient values can not be distinguished from hardware noise, or will need exponentially many measurements to do so. These challenges motivate the study of training strategies that avoid initialization on a barren plateau, as well as avoid creeping onto a plateau during the optimization process.

Indeed a number of different optimization strategies for PQCs have been explored, including deterministic and stochastic optimizers~\cite{Wecker2015,Romero2017quantum,nannicini2019performance,nakanishi2019sequential,zhou2018quantum,parrish2019jacobi,sung2020exploration}.
Regardless of the specific form of parameter update, the magnitudes of partial derivatives play a crucial role in descending towards the minimum of the objective function. Excessively small values will slow down the learning progress significantly, prevent progress, or even lead to false convergence of the algorithm to a sub-optimal objective function value. Crucially to this work, small values ultimately lead to a poor signal-to-noise ratio in PQC training algorithms due to the cost of information readout on a quantum device. Even if only sub-circuits of the overall circuit become sufficiently random during the training process, gradient values in a PQC will become exponentially small in the number of qubits~\cite{McClean2018}.

Moreover, in quantum-classical algorithms there is a fundamental readout complexity cost of $\mathcal{O}(1/\epsilon^\alpha)$~\cite{Knill2007}~as compared to a similar cost of $\mathcal{O}(\log(1/\epsilon^\alpha))$ classically. This is because classical arithmetic with floating point numbers for calculating analytic gradients may be done one digit at a time, incurring a cost $\mathcal{O}(\log(1/\epsilon^\alpha))$. In contrast, quantum methods with incoherent sampling, such as that utilized in NISQ algorithms, converge similarly to classical Monte Carlo sampling. This means that small gradient magnitudes can result in an exponential signal-to-noise problem when training quantum circuits. As a consequence, gradients become too small to be useful even for modest numbers of qubits and circuit depths, and a randomly initialized PQC will start the training procedure on a saddle point in the training landscape with no interesting directions in sight.

To utilize the capabilities of PQCs, methods that overcome this challenge have to be studied. Due to the vast success of gradient-based methods in the classical regime, this work is concerned with understanding how these methods can be adapted effectively for quantum circuits. We propose layerwise learning, where individual components of a circuit are added to the training routine successively. By starting the training routine in a shallow circuit configuration, we avoid the unfavorable type of random initialization described in \cite{McClean2018} and \cref{sec:bpp} which is inherent to randomly initialized circuits of even modest depth. In our approach, the circuit structure and number of parameters is successively grown while the circuit is trained, and randomization effects are contained to subsets of the parameters at all training steps. This does not only avoid initializing on a plateau, but also reduces the probability of creeping onto a plateau during training, e.g.\ when gradient values become smaller on approaching a local minimum. 

We compare our approach to a simple strategy to avoid initialization on a barren plateau, namely setting all parameters to zero, and show how the use of a layerwise learning strategy increases the probability of successfully training a PQC with restricted precision induced by shot noise by up to 40\% for classifying images of handwritten digits. Intuitively, this happens for reasons that are closely tied to the sampling complexity of gradients on quantum computers.  By restricting the training and randomization to a smaller number of circuit components, we focus the magnitude of the gradient into a small parameter manifold.  This avoids the randomization issue associated with barren plateaus, but importantly is also beneficial for a NISQ quantum cost model which must stochastically sample from the training data as well as the components of the gradient.  Simply put, with more magnitude in fewer components at each iteration, we receive meaningful training signal with fewer quantum circuit repetitions.

Another strategy to avoid barren plateaus was recently proposed by Grant et al.~\cite{Grant2019}, where only a small part of the circuit is initialized randomly, and the remaining parameters are chosen such that the circuit represents the identity operation. This prevents initialization on a plateau, but only does so for the first training step, and also trains a large number of parameters during the whole training routine. Another way to avoid plateaus was introduced in \cite{volkoff2020large}, where multiple parameters of the circuit are enforced to take the same value. This reduces the overall number of trained parameters and restricts optimization to a specific manifold, at the cost of requiring a deeper circuit for convergence \cite{Kiani2020}. 

In the classical setting, layerwise learning strategies have been shown to produce results comparable to training a full network with respect to error rate and time to convergence for classical NNs~\cite{Fahlman1990,Hettinger2017}. It has also been introduced as an efficient method to train deep belief networks (DBNs), which are generative models that consist of restricted Boltzmann machines (RBMs)~\cite{Hinton2006}. Here, multiple layers of RBMs are stacked and trained greedily layer-by-layer, where each layer is trained individually by taking the output of its preceding layer as the training data. In classical NNs, \cite{Bengio2007} shows that this strategy can be successfully used as a form of pre-training of the full network to avoid the problem of vanishing gradients caused by random initialization. In contrast to greedy layerwise pre-training, our approach does not necessarily train each layer individually, but successively grows the circuit to increase the number of parameters and therefore its representational capacity.

The remainder of this work is organized as follows: in \cref{sec:bpp} we give a short introduction to the barren plateau problem to motivate our training approach. \Cref{sec:layerwise} describes the layerwise learning strategy, which is then numerically demonstrated in \cref{sec:experiments} on the learning task of classifying handwritten digits.

\section{Barren plateaus in training landscapes}
\label{sec:bpp}

The notion of vast saddle points where first and higher order derivatives vanish, or barren plateaus, in quantum circuit training landscapes was formally introduced in \cite{McClean2018}. These plateaus result from basic features of the geometry of high-dimensional spaces. That work examines the gradients for quantum circuit training in a model known as random PQCs. For this model, parameter updates for gradient-based optimization of a PQC are calculated based on the expectation value of an observable measured on a state produced by a parametrized circuit and the circuits are drawn from a random distribution. The distribution of circuits amounts to choosing a set of gates uniformly at random, where some of the gates have continuous parameters amenable to differentiation. They show that a sufficient, but not necessary, condition for the vanishing of gradients with high probability is that for any division of the circuit into two pieces, the two pieces are statistically independent, and that one of them approximately matches the fully random distribution up to the second moment, or forms an approximate 2-design as described below. In other words, if one piece of the PQC forms an approximate 2-design, the whole circuit will be susceptible to exponentially vanishing gradients.

\begin{figure}
\captionsetup{justification=raggedright}
\includegraphics[width=0.5\textwidth]{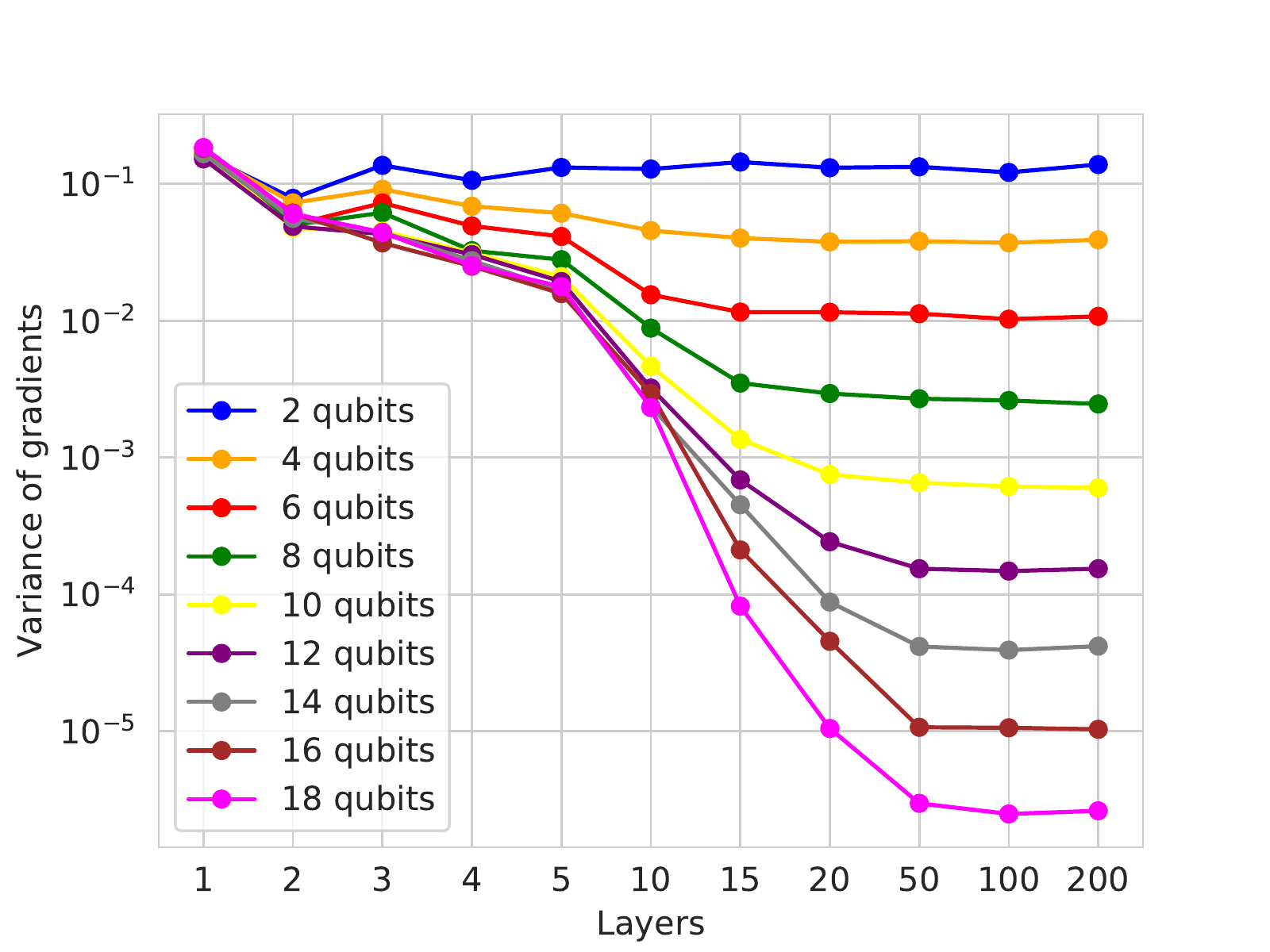}
\caption{Concentration of variance of gradients of the expectation value of the readout qubit. For random parametric quantum circuits, as circuits of different sizes converge to a 2-design, gradient values necessary for training vanish with increasing number of qubits and layers. Each data point is calculated over 1000 randomly generated circuits with all-to-all connectivity in each layer, and an initial layer of Hadamard gates on each qubit, to avoid a bias of gradients with respect to the circuit structure caused by initializing in an all-zero state.  Note that the average value of the gradient here is 0 in all cases. This serves as a reference point for the circuits we consider in later sections, and when our method is expected to be fighting problems related to concentration of the gradients to 0.}
\label{2-des_convergence}
\end{figure}

A particular class of random circuits that were studied numerically in \cite{McClean2018} were those which chose some discrete structure of gates determined by an underlying geometry (e.g.\ 1D line, 2D array, or all-to-all), and single qubit gates with a continuous parameter rotating around a randomly chosen axis. This assigns a parameter $\theta_i$ to each of these gates. To sample from the distribution of random circuits, each angle $\theta_i$ was drawn from the uniform distribution $\theta_i \in \mathcal{U}(0, 2\pi)$. Due to concentration of measure effects as described in \cite{McClean2018}, random PQCs with different sets of parameters will produce very similar outputs and their variance vanishes for sufficiently large circuits (those that reach 2-designs, as shown in \cref{2-des_convergence}).

Moreover, this represents a generic statement about the volume of quantum space for such a circuit where one expects to find trivial values for observables. 
In other words, sufficiently deep, arbitrary random PQCs will produce very similar expectation values regardless of the set of individual parameters. Consequently, the partial derivatives of an objective function based on expectation values of a random PQC will have extremely small mean and variance. These properties make them essentially untrainable for gradient-based or related local optimization methods on quantum devices, as the training landscapes in these scenarios are characterized by large regions where the gradient is almost zero, but which do not correspond to a local or global minimum of the objective function.

The origin of this phenomenon can be seen from the fact that a sufficiently deep sequence of random gates will yield an approximately Haar-distributed unitary. This results in the scenario described above, where expectation values concentrate to their average over the entire space. While executing a circuit that precisely outputs states from the Haar distribution takes time exponential in the number of qubits, there are efficient techniques to produce circuits that output states which approximate the first and second moment of the Haar distribution, which is formalized by the notion of $t$-designs for $t=2$. It has been shown in~\cite{Harrow2009} that random PQCs form approximate $2$-designs once they reach a certain depth, and \cite{McClean2018} shows how even a modest number of qubits and layers of random gates is enough for this. The depth of a quantum circuit required to reach this regime of barren plateaus depends on the number of qubits and allowed connectivity. It is thought that the depth required to reach a $2$-design scales roughly as $\mathcal{\tilde O}(n^{1/d})$ converging to a logarithmic required depth in the all-to-all connectivity limit~\cite{Boixo2018}, where $d$ is the dimension of the connectivity of the device and $n$ is the number of qubits.

These results demonstrate that randomly initialized PQCs of large depth become increasingly hard to train for any optimization method that can not avoid or escape plateaus during the optimization routine. \Cref{2-des_convergence} shows how the variance of partial derivatives obtained from circuits of different size and all-to-all connectivity in each layer exponentially decays as a function of the number of qubits and layers, as the circuits approach convergence to a 2-design. The variance of gradients was calculated over 1000 random circuit instances for each combination of number of qubits and number of layers, with an initial layer of Hadamard gates in each circuit to avoid a bias of gradients with respect to the circuit structure, caused by initializing in the all-zero state. With increasing number of qubits, convergence happens after a larger number of layers, yielding a variance that exponentially decays in the number of qubits. This decay is not only detrimental to learning, but more generally hinders extraction of meaningful values from quantum hardware, especially on near-term processors that are subject to noise.

\section{Layerwise learning}
\label{sec:layerwise}

In this section, we introduce layerwise learning (LL) for parametrized quantum circuits, a training strategy that creates an ansatz during optimization, and only trains subsets of parameters simultaneously to ensure a favorable signal-to-noise ratio. The algorithm consists of two phases.

\textit{Phase one}: The first phase of the algorithm constructs the ansatz by successively adding layers. The training starts by optimizing a circuit that consists of a small number $s$ of start layers, e.g.\ $s=2$, where all parameters are initialized as zero. We call these the initial layers $l_1(\Vec \theta_1)$:
\begin{equation}
    l_1(\Vec\theta_1) = \prod_{j=1}^{s}{U_{1_j}(\Vec\theta_{1_j})}W~,
\end{equation}
where $\Vec\theta_1$ is the set of parameters, containing one angle for each local rotation gate per qubit, and $W$ represents operators connecting qubits. After a fixed number of epochs, another set of layers is added, and the previous layers' parameters are frozen. We define one epoch as the set of iterations it takes the algorithm to see each training sample once, and one iteration as one update over all trainable parameters. E.g.\ an algorithm trained on 100 samples with a batch size of 20 will need 5 iterations to complete one epoch. The number of epochs per layer, $e_l$, is a tunable hyperparameter. Each consecutive layer $l_i(\theta_i)$ takes the form
\begin{equation}
    l_i(\Vec\theta_i) = U_i(\Vec\theta_i)W~,
\end{equation}
with a fixed $W$, as the connectivity of qubits stays the same during the whole training procedure. All angles in $\Vec\theta_i$ are set to zero when they are added, which provides additional degrees of freedom for the optimization routine without perturbing the current solution. The parameters added with each layer are optimized together with the existing set of parameters of previous layers in a configuration dependent on two hyperparameters $p$ and $q$. The hyperparameter $p$ determines how many layers are added in each step, and $q$ specifies after how many layers the previous layers' parameters are frozen. E.g.\ with $p=2$ and $q=4$, we add two layers in each step, and layers more than four back from the current layer are frozen. This process is repeated either until additional layers do not yield an improvement in objective function value, or until a desired depth is reached. The final circuit that consists of $L$ layers can then be represented by
\begin{equation}
    U(\Vec \theta) = \prod_{i=1}^L l_i(\Vec\theta_i)~.
\end{equation}

\textit{Phase two}: In the second phase of the algorithm, we take the pre-trained circuit acquired in phase one, and train larger contiguous partitions of layers at a time. The hyperparameter $r$ specifies the percentage of parameters that is trained in one step, e.g.\ a quarter or a half of the circuit's layers. The number of epochs for which these partitions are trained is also controlled by $e_l$. In this setting, we perform additional optimization sweeps where we alternate over training the specified subsets of parameters simultaneously, until the algorithm converges. This allows us to train larger partitions of the circuit at once, as the parameters from phase one provide a sufficiently non-random initialization. As the randomness is contained to shallower sub-circuits during the whole training routine, we also minimize the probability to creep onto a plateau during training as a consequence of stochastic or hardware noise present in the sampling procedure. 

\begin{figure}[t!]
\captionsetup{justification=raggedright}
\includegraphics[width=0.45\textwidth]{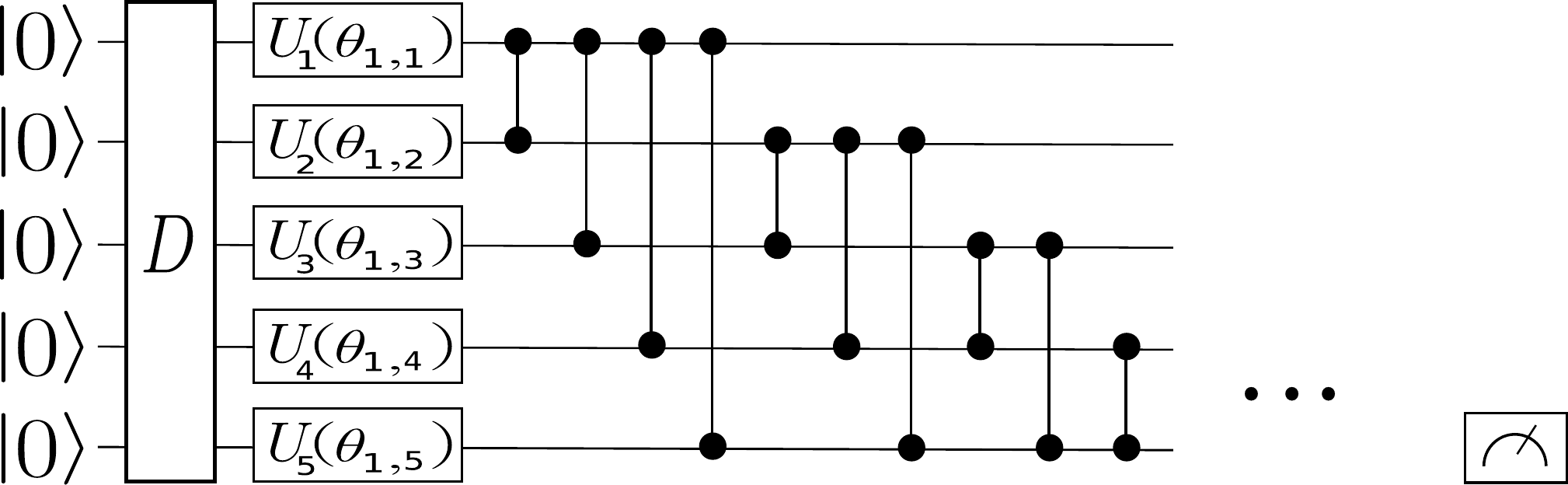}
\caption{Sample circuit layout of the first layer in an LL circuit. $D$ represents the data input which is only present once at the beginning of the circuit. The full circuit is built by successively stacking single rotation gates and two qubit gates to form all-to-all connected layers. For the classification example we show in \ref{sec:experiments}, a measurement gate is added on the last qubit after the last layer.}
\label{fig:all-to-all-layer}
\end{figure}

In general, the specific structure of layers $l_i(\Vec\theta_i)$ can be arbitrary, as long as they allow successively increasing the number of layers, e.g.\ like the hardware-efficient ansatz introduced in \cite{kandala2017hardware}. In this work, we indirectly compare the quality of gradients produced by our optimization strategy with respect to the results described in section \ref{sec:bpp} through overall optimization performance, so we consider circuits that consist of layers of randomly chosen gates as used in \cite{McClean2018}. They can be represented in the following form:
\begin{align} \label{eq:1}
    U(\Vec{\theta}) = \prod_{l=1}^{L}U_l(\Vec{\theta_l})W~,
\end{align}
where $U_l(\Vec \theta_l) = \prod_{i=1}^n \mathrm{exp}(-i \theta_{l,i} V_i)$ with a Hermitian operator $V_i$, $n$ is the number of qubits, and $W$ is a generic fixed unitary operator. For ease of exposition, we drop the subscripts of the individual gates in the remainder of this work. We consider single qubit generators $V$ which are the Pauli operators $X$, $Y$ and $Z$ for each qubit, parametrized by $\theta_l$, while $W$ are $CZ$ gates coupling arbitrary pairs of qubits. An example layer is depicted in \cref{fig:all-to-all-layer}.

The structure and parameters of a quantum circuit define which regions of an optimization landscape given by a certain objective function can be captured. As the number of parametrized non-commuting gates grows, this allows a more fine-grained representation of the optimization landscape~\cite{Kiani2020}. In a setting where arbitrarily small gradients do not pose a problem, e.g.\ noiseless simulation of PQC training, it is often preferable to simultaneously train all parameters in a circuit to make use of the full range of degree of freedom in the parameter landscape. We will refer to this training scheme as complete-depth learning (CDL) from now on. In a noiseless setting, LL and CDL will perform similarly w.r.t.\ the number of calls to a QPU until convergence and final accuracy of results, as we show in the appendix. 
This is due to a trade off between the number of parameters in a circuit and the number of sampling steps to convergence \cite{Kiani2020}. A circuit with more parameters will converge faster in number of training epochs, but will need more QPU calls to train the full set of parameters in each epoch. On the other hand, a circuit with fewer parameters will show a less steep learning curve, but will also need fewer calls to the QPU in each update step due to the reduced number of parameters. When we consider actual number of calls to a quantum device until convergence as a figure of merit, training the full circuit and performing LL will perform similarly in a noise-free setting for this reason. However, this is not true when we enter a more realistic regime, where measurement of gradient values will be affected by stochastic as well as hardware noise, as we will show on the example of shot noise in \cref{sec:experiments}.  In such more realistic settings, the layerwise strategy offers considerable advantage in time to solution and quality of solution.

As noted in the appendix of \cite{McClean2018}, the convergence of a circuit to a 2-design does not only depend on the number of qubits, their connectivity and the circuit depth, but also on the characteristics of the cost function used. This was further investigated in \cite{cerezo2020cost}, where cost functions are divided into those that are local and global, in the sense that a global cost function uses the combined output of all qubits (e.g.\ the fidelity of two states), whereas a local cost function compares values of individual qubits or subsets thereof (e.g.\ a majority vote). Both works show that for global cost functions, the variance of gradients decays more rapidly, and that barren plateaus will present themselves even in shallow circuits. As our training strategy relies on using larger gradient values in shallow circuit configurations, especially during the beginning of the training routine, we expect that LL will mostly yield an advantage in combination with local cost functions.

\section{Results}
\label{sec:experiments}

\subsection{Setup}
\label{sec:experimental_setup}

To examine the effectiveness of LL, we use it to train a circuit with fully-connected layers as described in \cref{sec:layerwise}. While fully-connected layers are not realistic on NISQ hardware, we choose this configuration for our numerical investigations because it leads circuits to converge to a 2-design with the smallest number of qubits and layers \cite{McClean2018}, which allows us to reduce the computational cost of our simulations while examining some of the most challenging situations. To compare the performance of LL and CDL we perform binary classification on the MNIST data set of handwritten digits, where the circuit learns to distinguish between the numbers six and nine. We use the binary cross-entropy as the training objective function, given by
\begin{equation}\label{eq:cross_entropy}
    -\mathcal{L}(\Vec \theta) = - \bigl(y \, \log{(E(\Vec \theta))} + (1-y) \log{(1-E(\Vec \theta))}\bigr)~,
\end{equation}
where $\log$ is the natural logarithm, $E(\Vec \theta)$ is given by a measurement in the $Z$-direction $M=Z_o$ on qubit $o$ which we rescale to lie between 0 and 1 instead of $-1$ and 1, $y$ is the correct label value for a given sample, and $\Vec \theta$ are the parameters of the PQC. The loss is computed as the average binary cross entropy over the batch of samples. In this case, the partial derivative of the loss function is given by
\begin{align}
    \frac{\partial \mathcal{L}(\Vec \theta) }{\partial \theta_i} = 
    y \frac{1}{E(\Vec \theta)} \frac{\partial E(\Vec \theta)}{\partial \theta_i} 
    - (1-y) \frac{1}{1 - E(\Vec \theta)} \frac{\partial E(\Vec \theta)}{\partial \theta_i}~.
\end{align}

To calculate the objective function value, we take the expectation value of the circuit of observable $M$,
\begin{equation}\label{eq:2}
    E(\Vec \theta) = \bra{\psi}U^{\dagger}(\Vec \theta) M U(\Vec \theta)\ket{\psi},
\end{equation}
where $\ket{\psi}$ is the initial state of the circuit given by the training data set. The objective function now takes the form $\mathcal{L}(E(\Vec \theta))$ and the partial derivative for parameter $\theta_i$ is defined using the chain rule as
\begin{equation}\label{eq:4}
    \frac{\partial \mathcal{L}}{\partial \theta_i} = \frac{\partial \mathcal{L}}{\partial E(\theta_i)} \cdot \frac{\partial E(\theta_i)}{\partial \theta_i}~.
\end{equation}

To compute gradients of $E(\Vec \theta)$, we use the parameter shift rule \cite{Schuld2019}, where the partial derivative of a quantum circuit is calculated by the difference of two expectation values as
\begin{multline}\label{eq:ps_rule}
	\frac{\partial E(\Vec \theta)}{\partial \theta_i} = r\Bigl(\bra{\psi} U^{\dagger}(\Vec \theta + s \hat \theta_i) M U(\Vec \theta + s\hat \theta_i) \ket{\psi} \\
	- \bra{\psi} U^{\dagger}(\Vec \theta -s \hat \theta_i) M U(\Vec \theta - s \hat \theta_i) \ket{\psi}\Bigr)~,
\end{multline}
where $\hat \theta_i$ is a unit vector in the direction of the $i$'th component of $\theta$, $s=\pi/4r$, and $r=0.5$.

We note that in the numerical implementation, care must be taken to avoid singularities in the training processes related to $E(\theta) = \{0, 1\}$ treated similarly for both the loss and its derivative (we clip values to lie in $[10^{-15}, 1-10^{-15}]$).  We choose the last qubit in the circuit as the readout $o$, as shown in \cref{fig:all-to-all-layer}. An expectation value of $0$ ($1$) denotes a classification result for class sixes (nines). As we perform binary classification, we encode the classification result into one measurement qubit for ease of implementation. This can be generalized to multi-label classification by encoding classification results into multiple qubits, by assigning the measurement of one observable to one data label. We use the Adam optimizer \cite{Kingma2015} with varying learning rates to calculate parameter updates and leave the rest of the Adam hyperparameters at their typical publication values. 

To feed training data into the PQC, we use qubit encoding in combination with principal component analysis (PCA), following \cite{Grant2018}. Due to the small circuits used in this work, we have to heavily downsample the MNIST images. For this, a PCA is run on the data set, and the number of principle components with highest variance corresponding to the number of qubits is used to encode the data into the PQC. This is done by scaling the component values to lie within $[0, 2\pi)$, and using the scaled values to parametrize a data layer consisting of local $X$-gates. In case of 10 qubits, this means that each image is represented by a vector $\Vec d$ with the 10 components, and the data layer can be written as $\prod_{i=1}^{10} \exp (-i d_i X_i)$.

Different circuits of the same size behave more and more similarly during training as they grow more random as a direct consequence of the results in \cite{McClean2018}. This means that we can pick a random circuit instance that, as a function of its number of qubits and layers, lies in the 2-design regime as shown in \cref{2-des_convergence}, and gather representative training statistics on this instance. As noted in \cref{sec:layerwise}, an LL scheme is more advantageous in a setting where training the full circuit is infeasible, therefore we pick a circuit with 8 qubits and 21 layers for our experiments, at which size the circuit is in this regime. When using only a subset of qubits in a circuit as readout, a randomly generated layer might not be able to significantly change its output. For example, if in our simple circuit in \cref{fig:all-to-all-layer}, $U_5(\theta_{1,5})$ is a rotation around the $Z$ axis followed only by $CZ$ gates, no change in $\theta_{1,5}$ will affect the measurement outcome on the bottom qubit. When choosing generators randomly from $\{X, Y, Z\}$ in this setting, there is a chance of $1/3$ to pick an unsuitable generator. To avoid this effect, we enforce at least one $X$ gate in each set of layers that is trained. For our experiments, we take one random circuit instance and perform LL and CDL with varying hyperparameters.

\subsection{Sampling requirements}
\label{sampling_requirements}

To give insight into the sampling requirements of our algorithm, we have to determine the components that we need to sample. Our training algorithm makes use of gradients of the objective function that are sampled from the circuit on the quantum computer via the parameter shift rule as described in \cref{sec:experimental_setup}.
The precision of our gradients now depends on the precision of the expectation values for the two parts of the r.h.s.\ in \cref{eq:ps_rule}. The estimation of an expectation value scales in the number of measurements $N$ as $\mathcal{O}(\frac{1}{\epsilon^{\alpha}})$, with error $\epsilon$ and $\alpha>1$~\cite{Knill2007}. For most near-term implementations using operator averaging, $\alpha=2$, resembling classical central limit theorem statistics of sampling. This means that the magnitude of partial derivatives $\frac{\partial E}{\partial \theta_i}$ of the objective function directly influences the number of samples needed by setting a lower bound on $\epsilon$, and hence the signal-to-noise ratio achievable for a fixed sampling cost. If all of the magnitudes of $\frac{\partial E}{\partial \theta_i}$ are much smaller than $\epsilon$, a gradient based algorithm will exhibit dynamics more resembling a random walk than optimization.

\subsection{Comparison to CDL strategies}
\label{sec:comparison}

We compare LL to a simple approach to avoid initialization on a barren plateau, which is to set all circuit parameters in a circuit to zero followed by a CDL training strategy. We argue that considering the sampling requirements of training PQCs as described in \cref{sampling_requirements}, an LL strategy will be more frugal in the number of samples it needs from the QPU. Shallow circuits produce gradients with larger magnitude as can be seen in \cref{2-des_convergence}, so the number of samples $1/\epsilon^2$ we need to achieve precision $\epsilon$ directly depends on the largest component in the gradient. This difference is exhibited naturally when considering the number of samples as a hyperparameter in improving time to solution for training. In this low sample regime, the training progress depends largely on the learning rate. A small batch size and low number of measurements will increase the variance of objective function values. This can be balanced by choosing a lower learning rate, at the cost of taking more optimization steps to reach the same objective function value. We argue that the CDL approach will need much smaller learning rates to compensate for smaller gradient values and the simultaneous update of all parameters in each training step, and therefore more samples from the QPU to reach similar objective function values as LL. We compare both approaches w.r.t.\ their probability to reach a given accuracy on the test set, and infer the number of repeated re-starts one would expect in a real-world experiment based on that.

In order to easily translate the results here to experimental impact, we also compute an average runtime by assuming a sampling rate of 10kHz. This value is assumed to be realistic in the near term future, based on current superconducting qubit experiments shown in \cite{arute2019quantum} which were done with a sampling rate of 5kHz, not including cloud latency effects. The cumulative number of individual measurements taken from a quantum device during training is defined as
\begin{equation}
    r_i = r_{i-1} + 2n_p m b~,
\end{equation}
where $n_p$ is the number of parameters (taken times two to account for the parameter shift rule shown in \cref{sec:experimental_setup}), $m$ the number of measurements taken from the quantum device for each expectation value estimation, and $b$ the batch size. This gives us a realistic estimate of the resources used by both approaches in an experimental setting on a quantum device.

\subsection{Numerical results}
\label{sec:numerical_results}

\begin{figure}
\captionsetup{justification=raggedright}
\includegraphics[width=0.5\textwidth]{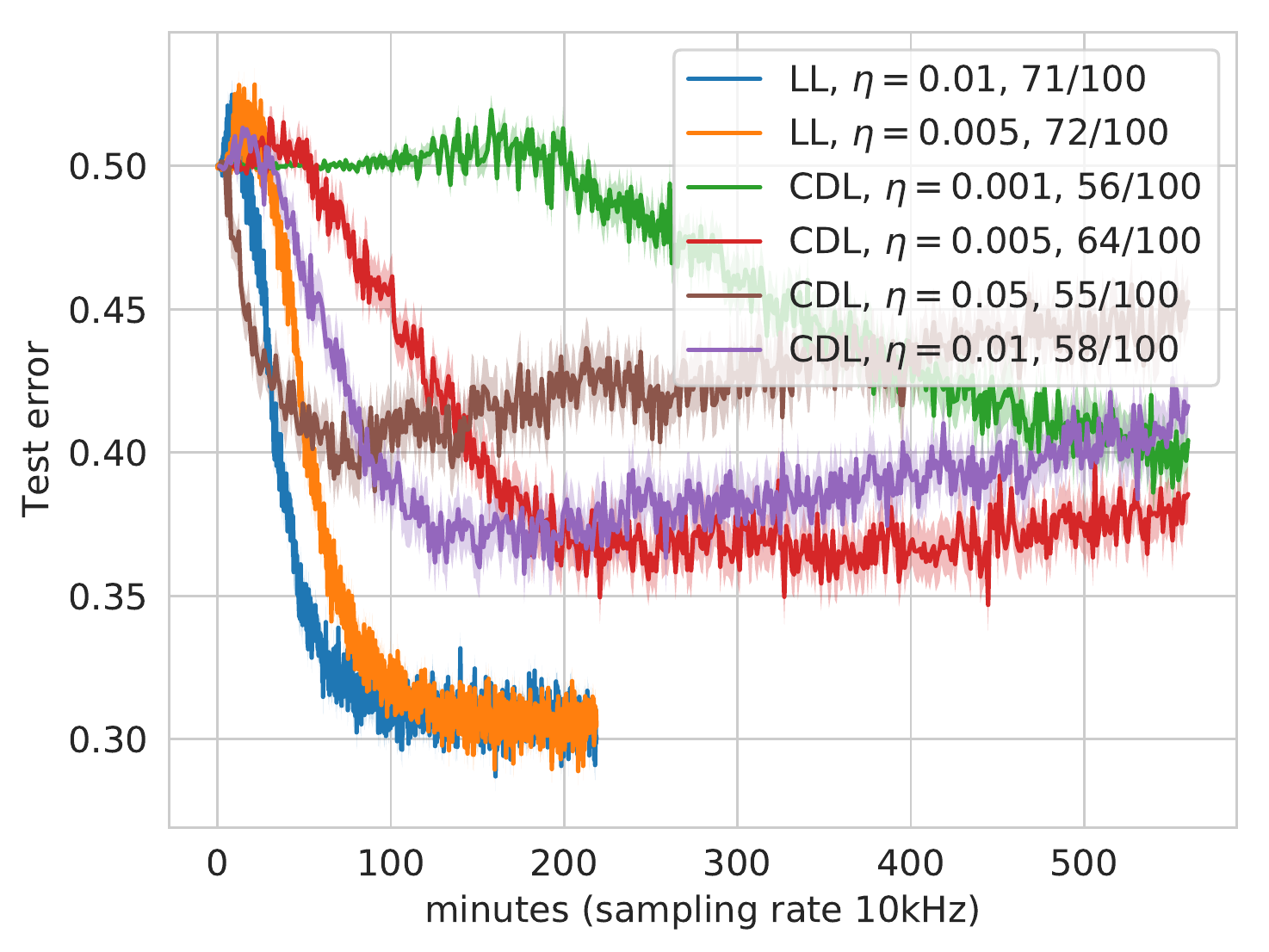}
\caption{Average test error as a function of runtimes for runs that have a final average test error less than 0.5 (random guessing) over the last ten training epochs, assuming a sampling rate of 10kHz and number of samples taken from the QPU computed as described in \cref{sec:comparison}. Numbers in labels indicate how many out of 100 runs were included in the average, i.e. fraction of runs that did not diverge in training, exhibiting less than 50\% error on the test set. Increasing test error for CDL runs with $\eta=0.01$ and $\eta=0.05$ is not due to overfitting, but due to a larger number of runs in the average that start creeping onto a plateau due to the increased learning rate.}
\label{fig:runtimes}
\end{figure}

For the following experiments, we use a circuit with 8 qubits, 1 initial layer and 20 added layers, which makes 21 layers in total. As can be seen in \cref{2-des_convergence}, this is a depth where a fully random circuit is expected to converge to a 2-design for the all-to-all connectivity that we chose. After doing a hyperparameter search over $p, q$ and $e_l$, we set the LL hyperparameters to $p=q=2$ and $e_l=10$, with one initial layer that is always active during training. This means that three layers are trained at once in phase one of the algorithm, and 10 and 11 layers are trained as one contiguous partition in phase two, respectively. For CDL, the same circuit is trained with all-zero initialization.

We argue that LL not only avoids initialization on a plateau, but is also less susceptible to randomization during training. In NISQ devices, this type of randomization is expected to come from two sources: (i) hardware noise, (ii) shot noise, or measurement uncertainty. The smaller the values we want to estimate and the less exact the measurements we can take from a QPU are, the more often we have to repeat them to get an accurate result. Here, we investigate the robustness of both methods to shot noise. The hyperparameters we can tune are the number of measurements $m$, batch size $b$ and learning rate $\eta$. The randomization of circuits during training can be reduced by choosing smaller learning rates to reduce the effect of each individual parameter update, at the cost of more epochs to convergence. Therefore we focus our hyperparameter search on the learning rate $\eta$, after fixing the batch size to $b=20$ and the number of measurements to $m=10$. This combination of $m$ and $b$ was chosen for a fixed, small $m$ after conducting a search over $b \in \{20, 50, 100\}$ for which both LL and CDL could perform successful runs that don't diverge during training. As we lower the batch size, we also increase the variance in objective function values similar to when the number of measurements is reduced, so these two values have to be tuned to match each other.  In the remainder of this section we show results for these hyperparameters, and different learning rates for both methods. All of the results are based on 100 runs of the same hyperparameter configurations. We use 50 samples of each class to calculate the cross entropy during training, and another 50 samples per class to calculate the test error. To compute the test error, we let the model predict binary class labels for each presented sample, where a prediction $\leq 0.5$ is interpreted as class 0 (sixes) and $> 0.5$ as class 1 (nines). The test error is then the average error over all classified samples. 

\begin{figure*}
  \captionsetup{justification=raggedright}
  \centering
  \subfloat[expected number of repetitions]{\includegraphics[scale=0.5]{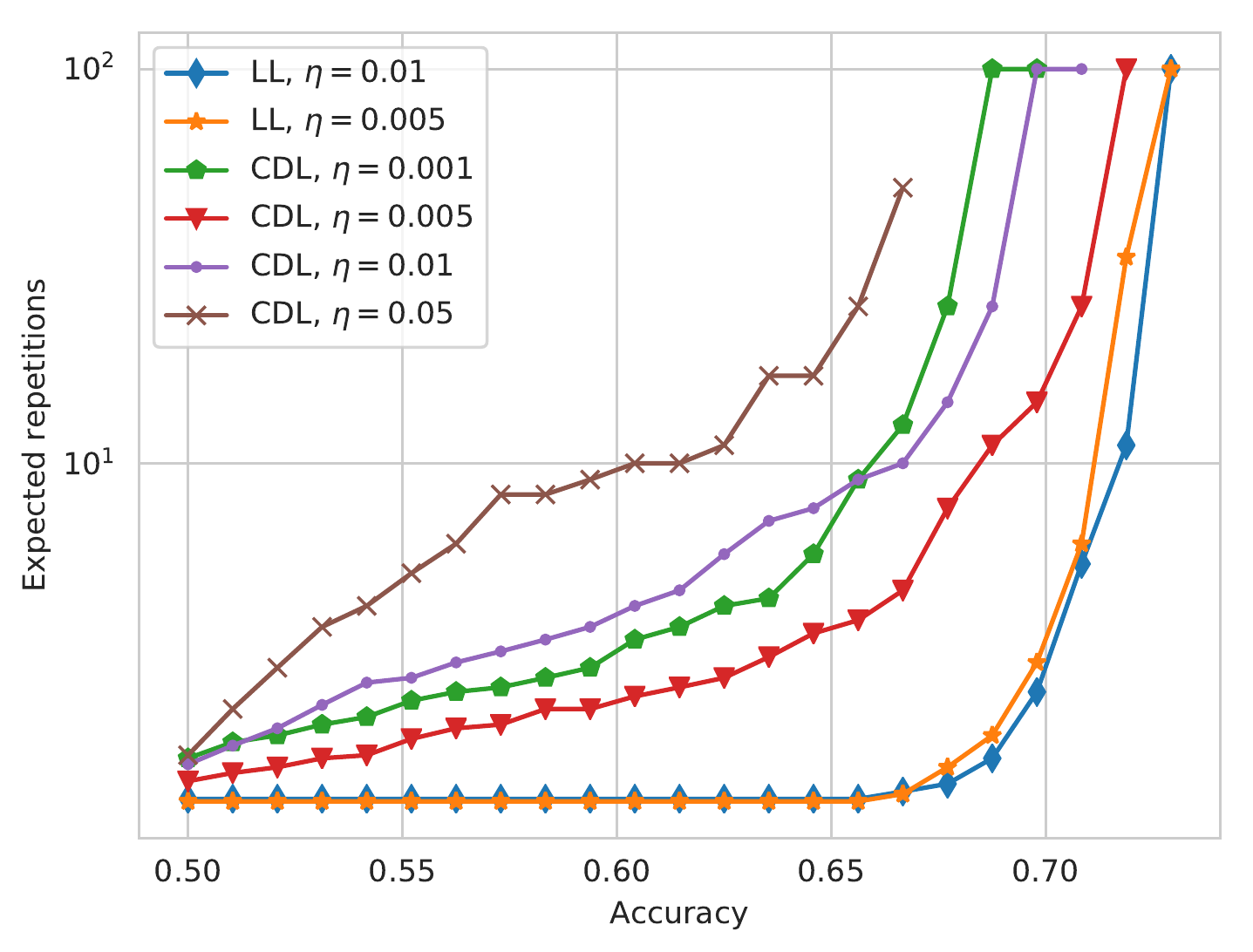}}\quad
  \subfloat[success probability]{\includegraphics[scale=0.5]{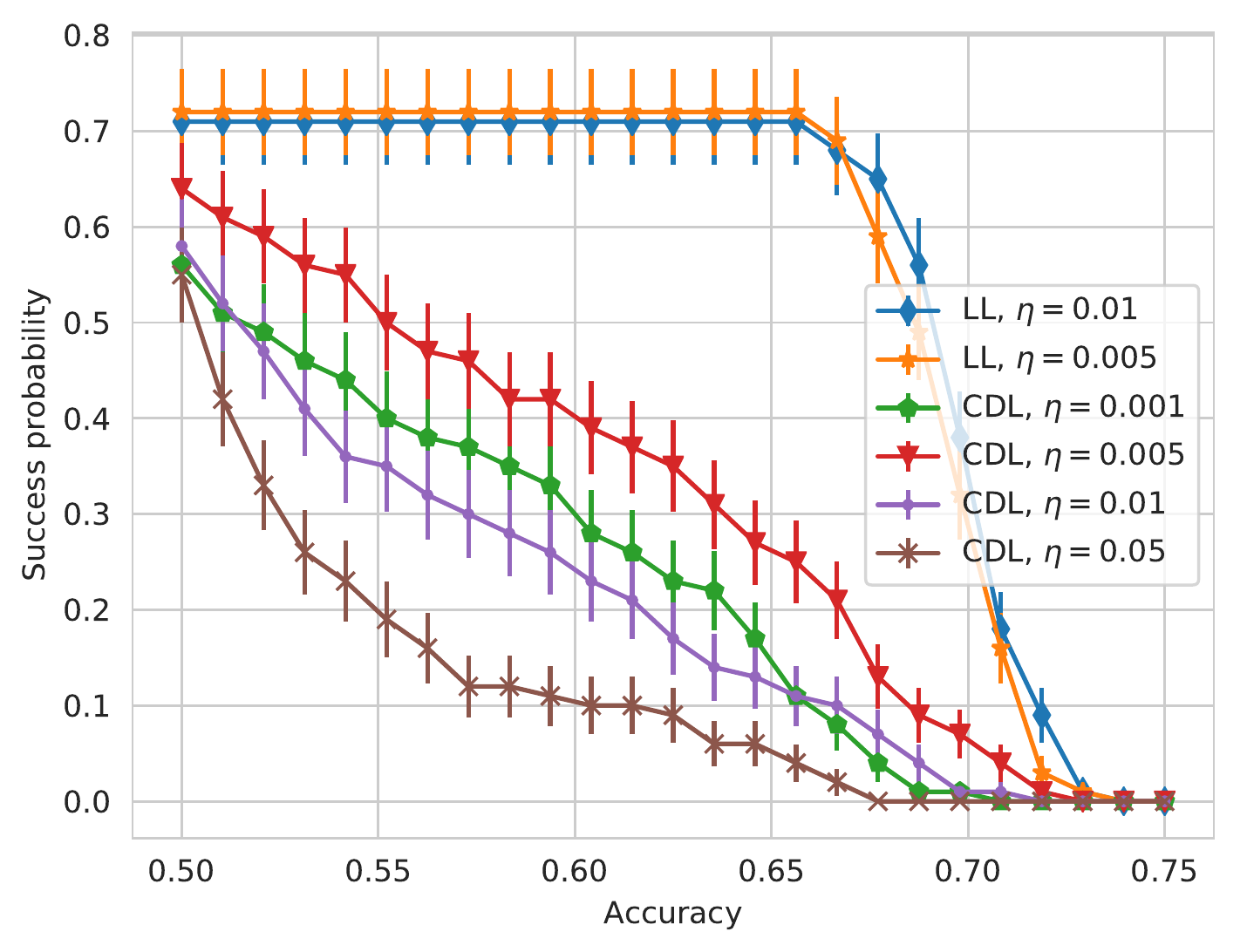}}
  \caption{LL decreases expected run time and increases probability of success on random restarts.  (a) Expected number of experiment repetitions needed until a given configuration reaches a certain accuracy defined as $(1 - \mathrm{error}_{\mathrm{test}})$, where $\mathrm{error}_{\mathrm{test}}$ is the average error on the test set, for LL and CDL with different learning rates. One experiment repetition constitutes in one complete training run of a circuit to a fixed number of epochs. Results are based on 100 runs for each configuration with $m=10$, $b=20$, and in case of LL, $e_l=10$. LL circuits better avoid randomization during training, and therefore need less than two repetitions on average for learning rates with varying magnitudes. CDL is more susceptible to entering a plateau during training in a noisy environment, as all parameters are affected on a perturbative update. This effect becomes more pronounced as learning rates are increased. (b) Probability of reaching a certain accuracy on the test set for the same configurations shown in (a). Success probability of LL stays constant up to an accuracy of 0.65 and starts decaying from there, as fewer runs reach higher accuracies on average. All CDL configurations have a lower success probability than the LL configurations overall, which decays almost linearly as we demand a higher average accuracy. Notably, the CDL configurations with highest success probability are also the ones with the highest runtime, as shown in \cref{fig:runtimes}.}
\label{fig:success_probability}
\end{figure*}

\Cref{fig:runtimes} shows average runtimes of LL and CDL runs that have a final average error on the test set that is less than 0.5, which corresponds to random guessing. We compute the runtime by computing the number of samples taken as shown in \cref{sec:comparison} and assume a sampling rate of 10kHz. Here, LL reaches a lower error on the test set on average, and also requires a lower runtime to get there. Compared to the CDL configuration with the highest success probability shown in \cref{fig:success_probability} (b) (red line), the best LL configuration (blue line) takes approximately half as much time to converge. This illustrates that LL does not only increase the probability of successful runs, but can also drastically reduce the runtime to train PQCs by only training a subset of all parameters at a given training step. Note also that the test error of CDL with $\eta=0.05$ and $\eta=0.01$ slowly increases at later training steps, which might look like overfitting at first. Here it is important to emphasize that these are averaged results, and what is slowly increasing is rather the percentage of circuits that have randomized or diverged at later training steps. The actual randomization in an individual run usually happens with a sudden jump in test error, after which the circuit can not return to a regular training routine anymore.

\Cref{fig:success_probability} (a) shows the number of expected training repetitions one has to perform to get a training run that reaches a given accuracy on the test set, where we define accuracy as $(1 - \mathrm{error}_{\mathrm{test}})$. One training run constitutes training the circuit to a fixed number of epochs, where the average training time for one run is shown in \cref{fig:runtimes}. An accuracy of 0.5 corresponds to random guessing, while an accuracy of around 0.73 is the highest accuracy any of the performed runs reached. We find that LL performs well for different magnitudes of learning rates as $\eta=0.01$ and $\eta=0.005$, and that these configurations have a number of expected repetitions that stays almost constant as we increase the desired accuracy. On average, one needs less than two restarts to get a successful training run when using LL. For CDL, the number of repetitions increases as we require the test error to reach lower values. The best configurations were those with $\eta=0.001$ and $\eta=0.005$, which reach similarly low test errors as LL, but need between 3 and 7 restarts to succeed in doing so. This is due to the effect of randomization during training, which is caused by the high variance in objective function values, and the simultaneous update of all parameters in each training step. In \cref{fig:success_probability} (b), we show the probability of each configuration shown in (a) to reach a given accuracy on the test set. All CDL configurations have a probability lower than 0.3 to reach an accuracy above 0.65, while LL reaches this accuracy with a probability of over 0.7 in both cases. This translates to the almost constant number of repetitions for LL runs in \cref{fig:success_probability} (a). Due to the small number of measurements and the low batch size, some of the runs performed for both methods fail to learn at all, which is why none of the configurations have a success probability of 1 for all runs to be better than random guessing.

\section{Conclusion and outlook}
\label{sec:conclusion}

We have shown that the effects of barren plateaus in QNN training landscapes can be dampened by avoiding Haar random initialization and randomization during training through layerwise learning. While performance of LL and CDL strategies is similar when considering noiseless simulation and exact analytical gradients, LL strategies outperform CDL training on average when experimentally realistic measurement strategies are considered. Intuitively, the advantage of this approach is drawn from both preventing excess randomization and concentrating the contributions of the training gradient into fewer, known components.  Doing so directly decreases the sample requirements for a favorable signal-to-noise ratio in training with stochastic quantum samples.  To quantify this in a cost effective manner for simulation, we reduce the number of measurements taken for estimating each expectation value. We show that LL can reach lower objective function values with this small number of measurements, while reducing the number of overall experiment repetitions until convergence to roughly half of the repetitions needed by CDL when comparing the configurations with highest success probability. This makes LL a more suitable approach for implementation on NISQ devices, where taking a large number of measurements is still costly and where results are further diluted by decoherence effects and other machine errors.  While our approach relies on manipulating the circuit structure itself to avoid initializing a circuit that forms a 2-design, it can be combined with approaches that seek to find a favorable initial set of parameters as shown in \cite{Grant2019}. The combination of these two approaches by choosing good initial parameters for new layers is especially interesting as the circuits grow in size. This work has also only explored the most basic training scheme of adding a new layer after a fixed number of epochs, which can still be improved by picking smarter criteria like only adding a new layer after the previous circuit configuration converged, or replacing gates in layers which provide little effect on changes of the objective function value similar to \cite{grimsley2019adaptive}.  Moreover, one could consider training strategies which group sets of coordinates rather than circuit layers. These possibilities provide interesting directions for additional research, and we leave their investigation for future works.

\section*{Acknowledgements}
AS, PS and ML acknowledge
funding from the European Union's Horizon 2020 research and innovation programme
under the Grant Agreement No. 828826.
AS and ML thank Michael Streif and Sheir Yarkoni for valuable discussions. AS thanks the TensorFlow Quantum team for providing early access to the library, which was used to perform the simulations in this work.

\bibliographystyle{apsrev4-1_with_title}
\bibliography{main}

\appendix

\section*{Appendix}\label{sec:appendix}

\begin{figure}[H]
\captionsetup{justification=raggedright}
\includegraphics[width=0.5\textwidth]{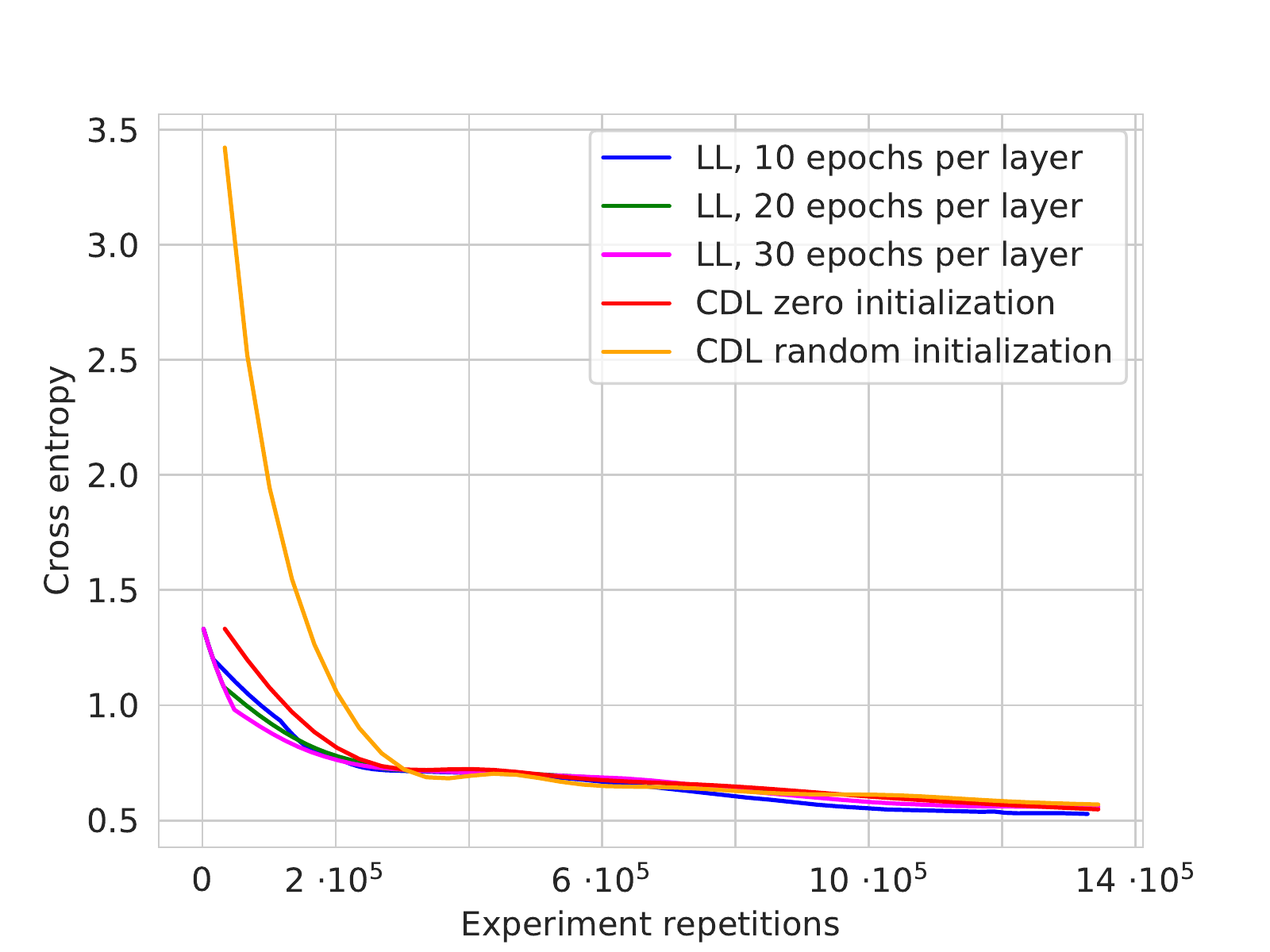}
\caption{Cross entropy of LL and CDL with exact gradient calculation corresponding to infinite number of measurements. When one assumes the unphysical situation of infinite measurements ($m=\infty$) all methods seem to perform similarly.  In particular, we compare LL to CDL with zero and random initialization, where the initial parameters for the latter are chosen uniformly from $[0, 2\pi)$. The hyperparameters for all configurations were set to $m=\infty$, $b=100$ and $\eta=0.01$. (For computing the number of experiment repetitions as defined in \cref{sec:comparison}, we drop $m$.)}
\label{fig:mnist_plot_exact}
\end{figure}

As alluded to in \cref{sec:layerwise}, LL and CDL perform similarly in a perfect simulation scenario, where we assume neither shot nor hardware noise. \Cref{fig:mnist_plot_exact} shows a comparison of LL and CDL under perfect conditions, i.e. infinite number of measurements and a batch size that corresponds to the number of samples, which enables computation of exact gradients. Here, the magnitude of gradients doesn't affect the learning process severely, as the Adam optimizer uses adaptive learning rates for each parameter and can therefore handle different ranges of gradient magnitudes well as long as there is some variance in the computed gradients. In this regime, both approaches show similar performance. 

\begin{figure}[H]
\captionsetup{justification=raggedright}
\includegraphics[width=0.5\textwidth]{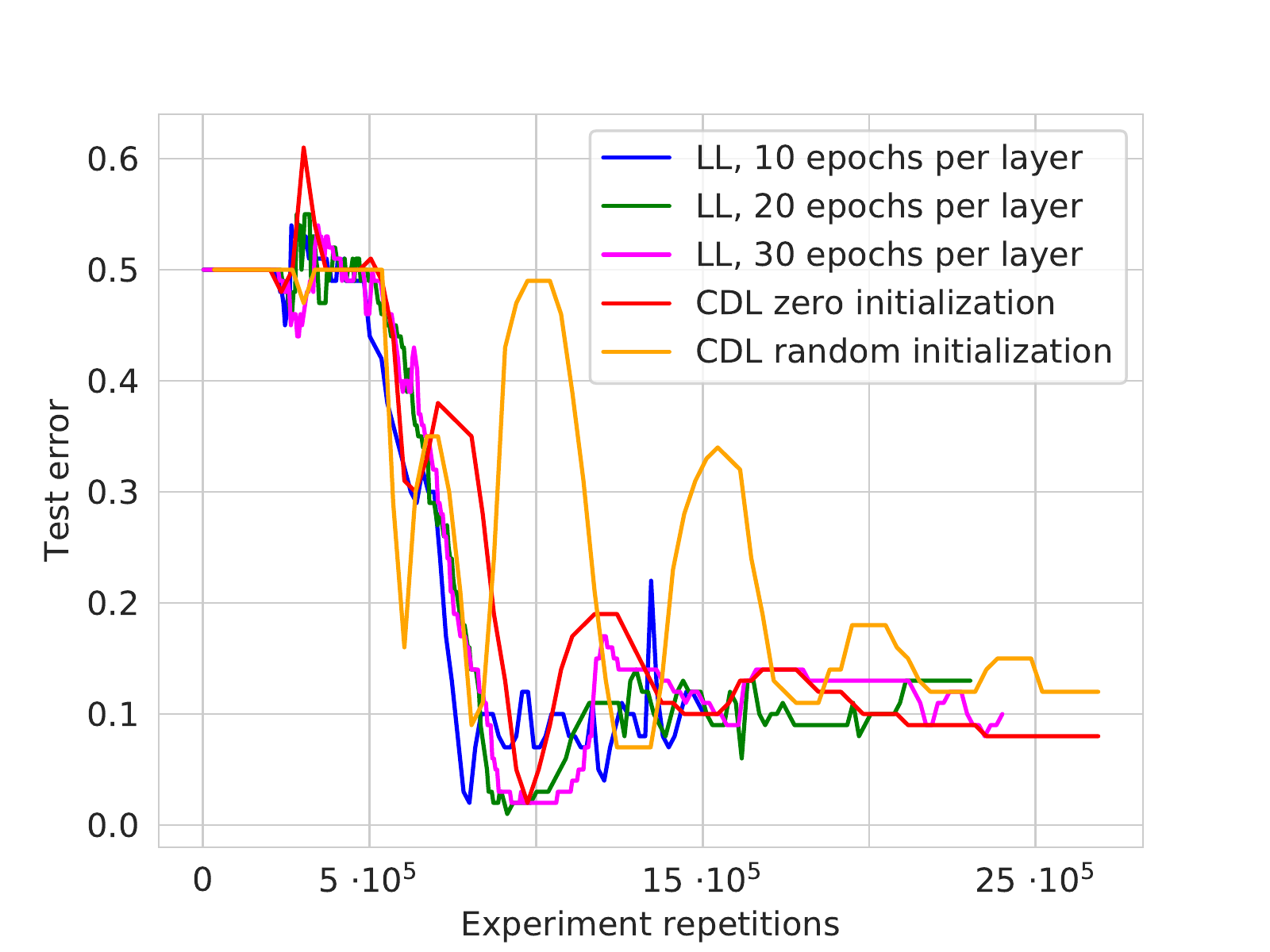}
\caption{Test error corresponding to the runs shown in \cref{fig:mnist_plot_exact}.  This further supports the observation that when one allows unphysical, arbitrary precision queries($m=\infty$), all tuned training strategies seem to perform similarly.}
\label{fig:mnist_plot_exact_te}
\end{figure}

The convergence rate of a PQC increases proportionally to the number of parameters in a model \cite{Kiani2020}, so the number of experiment repetitions is almost equal for LL and CDL. LL has less parameters and needs more epochs to converge due to this, whereas CDL needs more calls to the quantum device for one update step, but in turn needs less epochs to converge. In terms of cross entropy, both LL and CDL converge to a value of roughly $0.51$. The corresponding test error of all approaches, except for the randomly initialized CDL, reaches almost 0 but doesn't converge there and settles around an error of roughly $0.1$ eventually, as seen in \cref{fig:mnist_plot_exact_te}.

\end{document}